\documentclass[reprint,aps,prb,english,superscriptaddress,floatfix]{revtex4-2}

\usepackage{graphicx}
\usepackage{bm,amssymb,amsmath,mathrsfs,amstext,latexsym,physics}
\usepackage[T1]{fontenc}
\usepackage{booktabs}

\usepackage[usenames,dvipsnames]{xcolor}
\usepackage[colorlinks=true,citecolor=blue,urlcolor=blue,linkcolor=blue]{hyperref}
\usepackage[normalem]{ulem}

\usepackage{braket}

\newcommand{\be}{\begin{equation}}
\newcommand{\ee}{\end{equation}}
\def\ba{\begin{aligned}}
\def\ea{\end{aligned}}

\newcommand{\bea}{\begin{eqnarray}}
\newcommand{\eea}{\end{eqnarray}}

\begin{document}

\title{Semi-fractality and localization on a chiral Cayley tree}

\author{Carlo Vanoni}
\email{cvanoni@princeton.edu}
\affiliation{Department of Physics, Princeton University, Princeton, New Jersey, 08544, USA}
\author{Vladimir E. Kravtsov}
\email{kravtsov@ictp.it}
\affiliation{The Abdus Salam ICTP, Strada Costiera 11, 34151, Trieste, Italy}
\author{Aleksey V. Lunkin}
\affiliation{Nanocenter CENN, Jamova 39, 1000 Ljubljana, Slovenia}
\author{Boris L. Altshuler}
\affiliation{Department of Physics, Columbia University, 538 West 120th Street, New York, New York 10027, USA}

\date{\today}

\begin{abstract}
    We study a quantum particle hopping on an infinite Cayley tree with nearest-neighbor hopping amplitudes drawn from a distribution singular as $|t|^{-a}$ near weak links and no on-site disorder. Because the graph is bipartite, the model has chiral symmetry, which strongly affects the statistics of eigenstates at the center of the spectrum. We derive the two power-law tail exponents of the local-density-of-states distribution from the band-center cavity equation and compare the predictions with population dynamics. These tails imply an unusual form of wave-function statistics, which we call semi-fractality: the eigenstates occupy an extensive fraction of the system, but their higher moments behave as in a multifractal state. We find that the symmetry properties of the local-density-of-states distribution are not fixed only by the symmetry class, but vary continuously with the exponent controlling the power-law hopping distribution. As this exponent is changed, the system crosses from a semi-fractal regime to a localized one. At the transition, the wave functions realize an extreme intermediate form that we call semi-localized, simultaneously extended in their support but localized according to higher moments.
\end{abstract}

\maketitle

\section{Introduction}
\label{sec:intro}

A quantum state can require an extensive number of sites to carry its norm while retaining rare large amplitudes that dominate its higher moments.
This opens the possibility of quantum states that are spatially extended yet violate ergodicity.
Extended but nonergodic states have attracted considerable interest
in the study of disordered single-particle and many-body systems~\cite{anderson_absence_1958,abrahams1979scaling,evers2008anderson,altshuler1997quasiparticle,basko2006metal,oganesyan2007localization,abanin2019MBLcolloquium,sierant2025many,tikhonov2016anderson,tikhonov2021anderson,sarkar2026}. 

The old paradigm stating that the only non-ergodic phase is the one in which eigenfunctions are exponentially localized in real space (for a single-particle localization) or in the Hilbert space (for many-body localization) has changed in light of progress made recently. 
In fact, a handful of works demonstrated that the non-ergodic wave functions may be extended, exhibiting mono- or multi-fractal properties~\cite{biroli2012difference,de2014anderson,altshuler2016nonergodic,Kravtsov_2018,savitz2019anderson,zirnbauer2023wegner}. The corresponding measure that quantifies such properties is the set of fractal dimensions $D_{q}=dS_{q}/d\ln N$, where $N$ is the Hilbert space dimension and $S_{q}$ is the Rényi ($q\neq 1$) or Shannon-von Neumann ($q=1$) entropy associated with the wave function coefficients~\cite{janssen1994,rodriguez2011multifractal}. Note that $D_q$ is a non-increasing function of $q$.

In terms of the set of fractal dimensions $D_{q}$, the extended and non-ergodic phases can be classified as follows. The {\it localized phase} is such that $D_{1}=0$, implying that $D_{q}=0$ for all $q>1$. The {\it mono/multi-fractal phase} are such that $0<D_{1}<1$ (and thus $0<D_{q}<1$ for all $q>1$). The {\it ergodic phase} is defined by $D_{q}=1$ for {\it all} $q\geq 0$ (as happens, e.g., for the Porter-Thomas distribution).
The fractal dimension $D_{1}$ plays a special role in this classification, as it determines the support set of the wave function $\mathcal{S} \propto N^{D_{1}}$, which is the minimal number of sites that gives the normalization $\sum_{r\in\mathcal S}|\psi(r)|^2=1-\epsilon$ with any prescribed accuracy $\epsilon\ll 1$~\cite{deluca2013support}. The fractal dimensions can be used in place of the conductivity~\cite{abrahams1979scaling} to construct the scaling theory of Anderson localization on expander graphs~\cite{vanoni2024renormalization} and in high-dimensional lattices~\cite{altshuler2025,balducci2025}. 

However, the above classification does not cover all the possible options. The missing case is what we name {\it semi-fractal phase}, in which $D_{1}=1$ but $D_{q}<1$, for $q>\beta\geq 1$.
The support set of wave functions in such a phase is proportional to $N$, but other important measures, e.g., the return probability, $R(t)$, which is determined by the fractal dimension $D_{2}$, may show a typical multifractal behavior $R_{t}\propto t^{-D_{2}}$~\cite{chalker1988scaling,evers2008anderson,kravtsov2011}.

According to its definition, $D_{q}$ in the semi-fractal phase must be {\it non-analytic}, generally with a jump of the derivative $dD_{q}/dq$ at $q=\beta>1$. Such a jump may only happen when the spectrum of fractal dimensions $f(\alpha)$ has a linear segment with the slope $\beta>1$.
As $\beta$ tends to $1$ from above, the fractal dimension $D_{q}$ tends to zero for $q>1$, remaining equal to $1$ for $q\leq 1$. In this limit, the semi-fractal phase reaches its extreme form when $D_{q}$ jumps from $1$ to $0$ at $q=1$. Such a limiting case, which we call {\it semi-localized}, was discussed as a certain curiosity in Ref.~\cite{deluca2013support}, which did not have model or experimental realizations known at that time.

Recent studies of several different systems have reported behavior consistent with the semi-fractal statistics defined above.
One example is the experiment by Google Quantum AI on the distribution of wave function coefficients in the Hilbert space of qubits with XY interactions in two dimensions~\cite{lunkin2026hilbertspacesignaturesnonergodic}. At moderate and strong disorder, the measured distribution of $\ln|\psi|^2$ has a tail consistent with a linear segment of slope $\beta>1$ in $f(\alpha)$. The slope depends on disorder strength and approaches $\beta=1$ at very strong disorder.
Similar statistics were found within the cavity method for a tight-binding model on an Erd\H{o}s--R\'enyi graph with random hopping amplitudes~\cite{cugliandolo2024}. The authors attributed this behavior to clusters connected to the rest of the graph by a few links that may be very weak.
Another example is the $\beta$-ensemble of random matrices, whose level statistics coincide with that of classical particles with logarithmic interactions at an arbitrary temperature~\cite{dumitriu2002,das2025}.
An alternative mechanism for semi-fractality was identified by
Truong and Ossipov~\cite{truong2016} in the deformed Gaussian
unitary ensemble $H=W\widetilde{H}W$. In this nonchiral ensemble,
semi-fractality arises from a suitable deterministic power-law
profile imposed through the diagonal matrix $W$.

For the XY model, the Erd\H{o}s--R\'enyi graph, and the
$\beta$-ensemble, we argue that an exact or approximate chiral
structure is present, as discussed in Sec.~\ref{sec:discussion}.
These examples motivate the study of chiral models that exhibit semi-fractality. Chirality alone, however, is not sufficient: many chiral models exhibit neither semi-fractality nor semi-localization~\footnote{Additional conditions are needed for these phases to exist, and we leave the discussions of such conditions for future work.}. Here we illustrate this distinction with two examples.

The first is the exactly solvable problem of the $E=0$ states of a two-dimensional Dirac fermion in a random Abelian vector potential, which belongs to the chiral unitary symmetry class AIII~\cite{ludwig1994dirac,mudry1996cft,kogan1996liouville}. Its states are multifractal in the non-frozen regime and undergo freezing at strong disorder, but exhibit neither semi-fractality nor semi-localization as defined above~\cite{chamon1996localization,castillo1997exact}. Nevertheless, their typical fractal dimensions $D_q$ are non-analytic in $q$. At a disorder-dependent termination point $q_c$, the second derivative of $D_q$ is discontinuous in both the non-frozen ($q_c>1$) and frozen ($q_c<1$) regimes. At the freezing threshold $q_c=1$, the discontinuity is in the first derivative~\cite{chamon1996localization,castillo1997exact,carpentier2001glass,horovitz2002freezing}. Thus, non-analyticity of $D_q$ alone does not imply semi-fractality.

The second example is the one-dimensional model with deterministic long-range hopping $t_{nm}\sim|n-m|^{-s}$ studied in Ref.~\cite{Bahovadinov2026}. Its ground state is semi-localized in momentum space in the weak-disorder regime only for $1<s<3/2$. Here we denote the hopping exponent by $s$ to distinguish it from the exponent $a$ of the hopping distribution studied below. The ground state can be embedded as a zero-energy state of a chiral Hamiltonian through a supersymmetric (SUSY) construction, irrespective of $s$. The chiral structure therefore persists outside the range in which semi-localization occurs.

In this work, we study the bulk of an infinite Cayley tree with link disorder and no on-site disorder at $E=0$, using the cavity method and population dynamics~\cite{abou1973selfconsistent,Me87,parisi2019anderson,baroni2024corrections}, and compare with exact diagonalization on finite random regular graphs (RRGs).
The hopping amplitudes are drawn independently from a distribution
parametrized by $a$, which controls the abundance of weak links
and includes the Gaussian case at $a=0$.
Our Hamiltonian is therefore a random weighted adjacency matrix of a
prescribed graph, with independent and identically distributed
link amplitudes and no imposed site-dependent variance profile.
We therefore do not expect the deterministic deformation mechanism
of Ref.~\cite{truong2016} to apply for this model.
We restrict attention to Hermitian systems; note, however, that semi-fractality has also been found in the non-Hermitian skin effect on a Cayley tree~\cite{hamanaka2025}.

Our main finding is that in both systems (infinite Cayley tree and finite RRG) the distribution $P(\rho)$ of the local density of states (LDoS) $\rho(E;r)=\sum_{n}|\psi_{n}(r)|^{2}\,\delta(E-E_{n})$ has power-law tails at small and large $\rho$, with different exponents. 
We derive these exponents from the purely imaginary band-center Green's function and the positive cavity recursion in Sec.~\ref{sec:analytic_ldos}, and compare them with numerical estimates. 
These tails produce a linear segment in $f(\alpha)$ and a non-analytic dependence of $D_q$ on $q$, so that the extended band-center phase governed by these chiral cavity equations is semi-fractal. 
This behavior is already present for purely Gaussian hoppings ($a=0$); a nontrivial power-law dependence of the hopping distribution near the origin is therefore not required for semi-fractality. 
As the parameter $a$ that controls the abundance of weak links increases, a transition to a localized phase occurs. 
At the transition point $a=a_c$, the slope of $f(\alpha)$ reaches $\beta=1$ and $D_q$ jumps from $1$ to $0$ at $q=1$. 
This peculiar {\it half-ergodic, half-localized phase} is an extreme form of semi-fractality, first anticipated in Ref.~\cite{deluca2013support}. 

For this model, belonging to the chiral class BDI~\cite{altland2001,zirnbauer2006}, we also examine the inversion relation associated with the Mirlin-Fyodorov symmetry~\cite{mirlin2006}. 
The exponent $\gamma=\gamma(a)$ entering this relation varies continuously with the hopping-distribution parameter $a$ and is therefore not fixed by the symmetry class alone~\cite{gruzberg2011}. 
These band-center power laws are modified away from $E=0$, as discussed in the Supplemental Material~\cite{SupplementalMaterial}.

Finally, exact diagonalization reveals a complementary eigenstate-level signature of semi-fractality. When the site probabilities of an eigenstate are ordered by magnitude, their disorder-averaged profile displays a broad rank-dependent power law. This provides a possible microscopic interpretation of semi-fractality as arising from a hierarchy of eigenstate weights.

The paper is organized as follows. In Sec.~\ref{sec:model}, we introduce the chiral Cayley-tree model and relate finite-broadening LDoS moments to finite-size moments of wave-function amplitudes. In Sec.~\ref{sec:LDOS_distr}, we keep the two tail exponents as parameters and derive the corresponding eigenfunction statistics. Section~\ref{sec:analytic_ldos} determines these exponents analytically from the band-center cavity equation. In Sec.~\ref{sec:population_dyn}, we present the population-dynamics results and compare the fitted exponents with these predictions. Section~\ref{sec:theory_num} compares the LDoS and eigenfunction statistics with exact diagonalization on finite random regular graphs, and examines the rank-ordered eigenstate probabilities. We conclude in Sec.~\ref{sec:discussion}.

\section{Model and formalism for the chiral orthogonal symmetry class}
\label{sec:model}

As anticipated in the Introduction, we consider the problem of a quantum particle hopping on a Cayley tree with branching number $k=2$, described by the Hamiltonian
\begin{equation}
    H = \sum_{\langle i,j \rangle} t_{ij} (c^{\dagger}_i c_j + \mathrm{h.c.})
\end{equation}
where $c^{\dagger}_i$ and $c_i$ are the creation and annihilation operators at site $i$, the sum runs over nearest-neighbor sites on the tree (denoted by $\langle \cdot,\cdot \rangle$), and the random numbers $t_{ij} = t_{ji}$ are distributed according to the distribution
\begin{equation}\label{eq:p_t}
    p(t) = \frac{1}{\Gamma \left( \frac{1-a}{2}\right)}\frac{e^{-t^{2}}}{|t|^{a}}, \quad (a<1),
\end{equation}
which reduces to a Gaussian for $a = 0$. The amplitudes on distinct links are independent and identically distributed; no site-dependent variance profile is imposed. The parameter $a$ controls the abundance of weak links: $p(t)$ vanishes at the origin for $a<0$, is finite there for $a=0$, and diverges for $0<a<1$. A Cayley tree is bipartite at any size, meaning that it can be divided into two sublattices such that a site from one sublattice is linked only to sites belonging to the other. The bipartite nature of the graph endows the problem with a chiral symmetry preserving both the time-reversal and particle-hole symmetries with ${\cal T}^{2}=+1$ and ${\cal P}^{2}=+1$, respectively. This means, in particular, that each eigenstate with the eigenvalue $E_{n}$ and eigenfunction $\psi_{E_n}(r)$ has a counterpart with the eigenvalue $-E_{n}$ and eigenfunction having opposite sign in one of the two sublattices (therefore, $|\psi_{E_n}(r)|^2 = |\psi_{-E_n}(r)|^2$).

Consequently, the single-point retarded/advanced Green's functions $G^{R/A}(E;r,r;\eta)\equiv G^{R/A}$ at $E=0$ obey the symmetry:
\begin{align}
G^{R/A}&=\sum_{n} \frac{|\psi_{n}|^{2}}{-E_{n}\pm i \eta} = \sum_{n} \frac{|\psi_{n}|^{2}}{+E_{n}\pm i \eta}\\
&= \mp i \eta \sum_{n} \frac{|\psi_{n}|^{2}}{E_{n}^{2}+ \eta^{2}},
\end{align}
i.e., the Green's functions $G^{R/A}$ ($G^{R}={G^{A}}^{*}$) are pure imaginary.

In the remainder of this section, we derive the relation between the moments of the LDoS at finite broadening $\eta$ in the infinite system and the moments of wave-function amplitudes in a finite Hermitian system. This relation provides the bridge from the LDoS distribution $P(\rho)$ to the multifractal spectrum of eigenfunctions: once the tail exponents of $P(\rho)$ determine the scaling of $\langle \rho^q\rangle$, Eq.~\eqref{final} gives the corresponding scaling of $I_q=N\langle |\psi|^{2q}\rangle$, from which $D_q$ and $f(\alpha)$ follow. 
We start by considering $G^{R}=-i\pi\rho$, where $\rho$ is the local density of states (LDoS). We have at $E=0$:
\begin{align}\label{rho_to_q}
\rho^{q}&= \left (\frac{\eta}{\pi}\right)^{q}\,\left(\sum_{n} \frac{|\psi_{n}|^{2}}{E_{n}^{2}+ \eta^{2}} \right)^{q}\nonumber\\
&= \left (\frac{\eta}{\pi}\right)^{q}\,\sum_{n_{1}...n_{q}}\prod_{i=1}^q\frac{|\psi_{n_{i}}|^{2}}{E_{n_{i}}^{2}+ \eta^{2}}\nonumber\\
&\approx
 \left(\frac{\eta}{\pi}\right)^{q}\,\sum_{n} \frac{|\psi_{n}|^{2q}}{(E_{n}^{2}+ \eta^{2})^{q}}.
\end{align}
For integer $q$ in the r.h.s., the most divergent terms with all $n_{i}$ ($i=1,2,...q$) equal to each other make the main contribution in the limit $\eta \to 0$. We now introduce:
 \begin{equation}\label{rho_q}
\rho_{q}(\xi)=N^{-1}\sum_{n}|\psi|^{2q}\,\delta(\xi-E_{n}).
\end{equation}
Then the summation in Eq.~\eqref{rho_to_q} can be replaced by integration. After averaging over disorder, we obtain:
\begin{equation}\label{main}
\langle \rho^{q}\rangle =N\left(\frac{\eta}{\pi}\right)^{q}\int_{-\infty}^{+\infty}d\xi\,\frac{\langle \rho_{q}(\xi)\rangle}{(\xi^{2}+\eta^{2})^{q}}.
\end{equation}
The $\delta$-function in Eq.~\eqref{rho_q} should be regularized, e.g.:
\begin{equation}
\delta(x)\rightarrow \delta_{\eta}(x)=\left\{\begin{matrix}1/(2\eta), & |x|<\eta\cr 0, & {\rm otherwise}  \end{matrix}  \right.
\end{equation}
To make contact with the exact diagonalization calculations, in which only the level closest to $E=0$ is taken into account, we have to choose $\eta=\eta_{c}$ such that inside the energy box of width $\eta_{c}$ there are {\it one or a few} levels. The condition for that is:
\begin{equation}
\eta_{c}\,N\bar\rho(\eta_{c})\sim 1,
\end{equation}
where $\bar\rho(E)$ is the mean DoS:
\begin{equation} 
\bar\rho(\eta)=\left\langle N^{-1}\sum_{n}\delta_{\eta}(E_{n})\right\rangle
\end{equation}
Assuming that the box-regularized mean DoS and the Lorentzian-regularized mean LDoS have the same scaling with $\eta$\footnote{For the same broadening kernel, the spatially averaged LDoS equals the DoS exactly, by eigenfunction normalization. No statistical independence between amplitudes and level fluctuations is required. Here the two kernels are assumed to give the same leading scaling near $E=0$.}, we obtain a condition for $\eta$ that should be found self-consistently:
\begin{equation}\label{eta_c}
\eta_{c}\,N\,\langle\rho\rangle|_{\eta=\eta_{c}}\sim 1.
\end{equation}
The solution of the above condition gives the value of $\eta=\eta_{c}$ that has to be substituted into the $\eta$-dependent moments of the LDoS obtained from population dynamics to convert the distribution of the LDoS at $N=\infty$ and a finite $\eta$ into the distribution of $\psi^{2}$ at $\eta=0$ and a finite $N$.

Under the condition in Eq.~\eqref{eta_c}, Eq.~\eqref{rho_q} provides the following relationship between the moment of LDoS and that of the wave function amplitude:
\begin{equation}\label{rho_q-psi_q}
\langle \rho_{q}\rangle = (N\eta_{c})^{-1}\,\langle |\psi|^{2q}\rangle.
\end{equation}
Finally, from Eq.~\eqref{main} we obtain the desired relation:
\begin{equation}\label{final}
I_{q}\equiv N\langle|\psi|^{2q} \rangle\sim (N\eta_{c})\,\eta_{c}^{q-1}\,\langle \rho^{q}\rangle|_{\eta=\eta_{c}},
\end{equation}
where $\eta_{c}$ is found from Eq.~\eqref{eta_c}.

\section{Distribution of the LDoS and eigenfunction statistics}
\label{sec:LDOS_distr}

We first keep the two LDoS tail exponents as parameters and derive their consequences for eigenfunction statistics. Their values are obtained from the band-center cavity equation in Sec.~\ref{sec:analytic_ldos} and compared with population dynamics in Sec.~\ref{sec:population_dyn}. We write the two regimes as
\begin{equation}\label{P_rho}
P(\rho)\sim 
\begin{cases}
    \rho^{1+\beta-\gamma}, & \eta\ll \rho\ll 1 \cr
    \rho^{-(1+\beta)}, & \eta^{-1}\gg \rho\gg 1.
\end{cases}
\end{equation}
It is easy to check that at $\beta>0$ and $\beta>\gamma-2$ the normalization factor $c^{-1}=\int_{\eta}^{\eta^{-1}}P(\rho)\,d\rho\sim 1$ in $P(\rho)$ of the form Eq.~\eqref{P_rho} is independent of $\eta\rightarrow +0$. 
\begin{figure}
    \centering
    \includegraphics[width=0.8\linewidth]{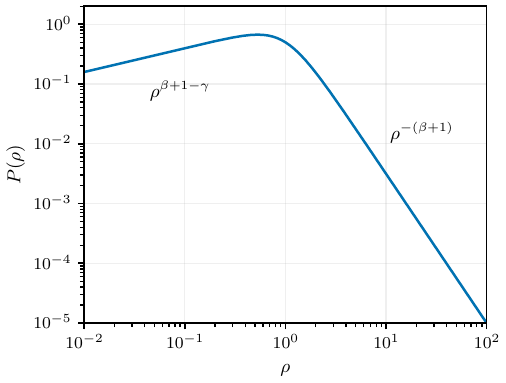}
    \caption{Schematic representation of the two power-law regimes of the LDoS distribution, derived in Sec.~\ref{sec:analytic_ldos} and observed in population dynamics (Sec.~\ref{sec:population_dyn}).} The small-$\rho$ regime is described by a power-law $\rho^{\beta +1 -\gamma}$, while the large-$\rho$ regime is described by $\rho^{-(\beta+1)}$.
\label{fig:D_q-f_a}
\end{figure}

\begin{figure}
\centering
\includegraphics[width=0.8\linewidth]{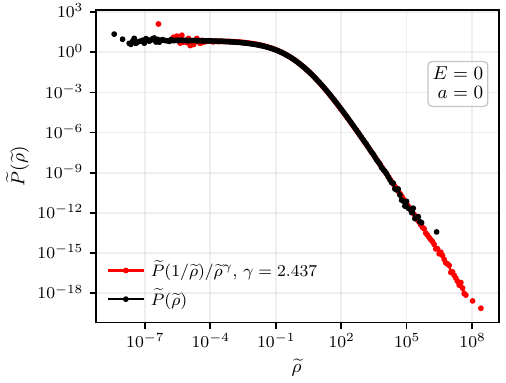}
    \caption{Comparison between $\widetilde P(\widetilde\rho)$ (black) and
    $\widetilde P(1/\widetilde\rho)/\widetilde\rho^{\gamma}$ (red) for
$a=0$ and $E=0$, using the fitted value $\gamma=2.44$. The asymptotic tail prediction is $\gamma=2.5$ in Eq.~\eqref{eq:analytic_model_exponents}.}
\label{fig:MF_symmetry}
\end{figure}

The parameter $\beta$ controls the large-$\rho$ tail, while $\gamma$ relates its exponent to that of the small-$\rho$ tail. A stronger statement is the inversion relation---usually known as Mirlin-Fyodorov symmetry~\cite{mirlin1994distribution}---for the full rescaled distribution $\tilde{P}(\tilde{\rho})=P(\rho)\rho_{0}$,
\begin{equation}\label{sym}
\tilde{P}(1/\tilde{\rho})=\tilde{\rho}^{\gamma}\,\tilde{P}(\tilde{\rho}),
\end{equation}
where $\tilde{\rho}=\rho/\rho_0$. Figure~\ref{fig:MF_symmetry} tests this relation numerically for Gaussian hoppings. The analytical derivation of the tail exponents below does not establish Eq.~\eqref{sym} for the full distribution. If Eq.~\eqref{sym} holds, normalization gives the following condition on its center $\rho_0$:
\begin{equation}
    \rho_{0}^{\gamma-2}= \langle \rho^{\gamma-2}\rangle.
\end{equation}
For all three Dyson symmetry classes, $\gamma=3$. It also takes a universal (independent of the further details of the system) value for the class CI ($\gamma=4$) and the class C ($\gamma=5$). However, for the {\it chiral symmetry classes} studied here, the parameter $\gamma$ may depend on system details and can even be a continuous function of the disorder strength parameter \cite{gruzberg2011}. Indeed, population dynamics shows that both $\gamma=\gamma(a)$ and $\beta=\beta(a)$ depend on $a$ in our model. However, the relations  
\begin{equation}
\gamma-2<\beta,\qquad\beta>0
\end{equation}
remain true for all values of $a<1$. 

At finite observation energy $E$, the LDoS distribution departs from its
chiral band-center form at $\rho\sim1/|E|$. The crossover and the effect
of on-site disorder are discussed in the Supplemental Material~\cite{SupplementalMaterial}.

For $q\geq0$ and $q\ne\beta$, integrating $\rho^{q}$ with the distribution function $P(\rho)$, Eq.~\eqref{P_rho}, gives~\cite{cugliandolo2024}
\begin{equation}\label{mom_rho_q}
\langle \rho^{q}\rangle \sim \left\{\begin{matrix} \eta^{0}, & q<\beta\cr \eta^{\beta-q},& q>\beta \end{matrix} \right.
\end{equation}
At $q=\beta$, the moment grows as $\ln(1/\eta)$.
We observe that, for $q=1$,
\begin{equation}
\langle \rho\rangle \sim \left\{\begin{matrix} O(1),& \beta>1\cr
\eta^{\beta-1}, &\beta<1\end{matrix}\right.
\end{equation}
We see that the cases $\beta>1$ and $\beta<1$ are fundamentally different. In the former case, the mean LDoS (and the mean DoS) is finite in the limit $\eta\rightarrow 0$, while in the latter it is divergent at $E=0$. Such zero-energy singularities are a familiar feature of chiral random hopping problems and their field-theory descriptions~\cite{mudry2003}. We will consider these two cases separately.
\begin{figure}
\centering
\includegraphics[width=0.8\linewidth]{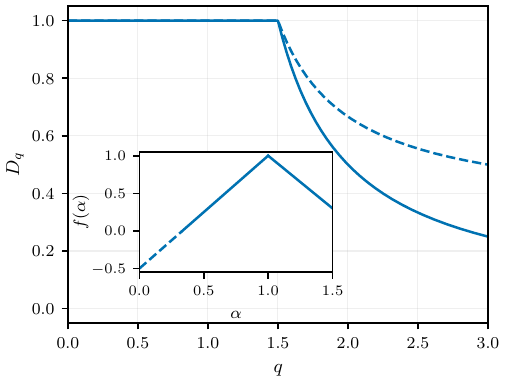}
\includegraphics[width=0.8\linewidth]{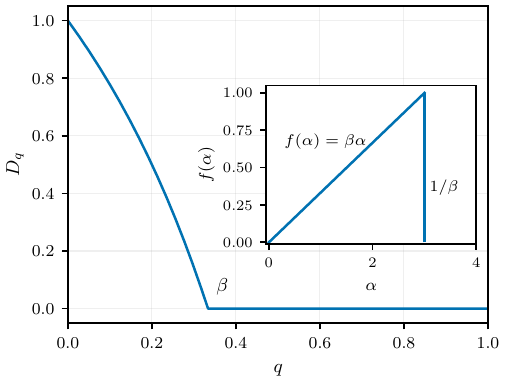}
    \caption{{\it Top panel}: Example of a semi-fractal $D_{q}$ with $\beta=1.5$. The solid line corresponds to $D_{q}$ with the rare events with large wave function amplitudes (shown by the dashed line in the inset) taken into account, while the dashed line corresponds to $D_{q}$ deduced from the typical average of the moments. In the inset, the corresponding spectrum of fractal dimensions is shown, which has a linear segment $f(\alpha)=\beta(\alpha-1)+1$ with the slope $\beta=1.5$ for $\alpha<1$ and an opposite linear segment of slope $\gamma-\beta-2=-1.4$, fixed by the small-$\rho$ tail with $\gamma=2.1$. {\it Bottom panel}: $D_{q}$ for $\beta<1$, corresponding to the localized case. The inset shows the corresponding $f(\alpha)$.}
\label{fig:beta_smaller_one}
\end{figure}
\subsection{Case $\beta>1$.} 
\label{sec:beta>1}
In this case, Eq.~\eqref{eta_c} gives a usual $\eta_{c}\sim N^{-1}$, and thus from Eq.~\eqref{final} we obtain:
\begin{equation}
    I_{q} \sim  \left\{\begin{matrix} N^{1-q}, & 0<q<\beta \cr N^{1-\beta}, & q>\beta>1 \end{matrix} \right. .
\end{equation}
The fractal dimension $D_{q}$ is then given by \cite{cugliandolo2024}:
\begin{equation}\label{D_q}
    D_{q}=-\frac{d\ln I_{q}}{d\ln N}\frac{1}{q-1}=\left\{\begin{matrix}1, & q<\beta \cr
    \frac{\beta-1}{q-1}<1, & q>\beta>1 \end{matrix} \right..
\end{equation}
Thus, $D_1=1$ and the support set is extensive, while $D_q<1$ for $q>\beta$ [see Fig.~\ref{fig:beta_smaller_one}]. This is the semi-fractal phase defined in the Introduction. In particular, $D_2<1$ when $1<\beta<2$. As $\beta\to1^+$, the dimensions at fixed $q>1$ tend to zero, whereas those at fixed $0\leq q<1$ remain equal to one. We call this limiting spectrum {\it semi-localized}.

\subsection{Case $\beta<1$.}
\label{sec:beta<1}
Now consider the case $\beta<1$. In this case, $\eta_{c}$ should be found self-consistently from Eq.~\eqref{eta_c}. Using $\langle \rho\rangle \sim \eta^{\beta-1}$ we obtain:
\begin{equation}
\eta_{c}\sim N^{-\frac{1}{\beta}}.
\end{equation}
Therefore, Eq.~\eqref{final} gives:
\begin{equation}
I_{q}=N\langle |\psi|^{2q}\rangle\sim \left\{\begin{matrix} N^{1-\frac{q}{\beta}},& q<\beta \cr
O(1), & q>\beta\end{matrix}\right.
\end{equation}
This corresponds to the localized state [see Fig.~\ref{fig:beta_smaller_one}]:
\begin{equation}\label{D_q_beta_less_1}
D_{q}=\left\{\begin{matrix} 1-\frac{q}{1-q}\frac{1-\beta}{\beta},& q<\beta<1\cr
0, & q>\beta \end{matrix} \right.
\end{equation}
since $D_1 = 0$.


\section{Analytical derivation of the LDoS tail exponents}
\label{sec:analytic_ldos}

We now determine the two tail exponents left as parameters in Sec.~\ref{sec:LDOS_distr}. The derivation uses the band-center cavity equation; the numerical comparison follows in Sec.~\ref{sec:population_dyn}. 
At the band center, chiral symmetry makes the diagonal and cavity Green's functions purely imaginary: the paired states at $E$ and $-E$ have equal local weights, so their real contributions cancel. Writing $G_{\mathrm{cav}}^R=-ix$ and $G_{ii}^R=-i\pi\rho$, with $x,\rho>0$, the cavity equation becomes
\begin{equation}
    x\overset{d}{=}\left(\eta+\sum_{j=1}^{k}\tau_jx_j\right)^{-1},
    \qquad
    \pi\rho\overset{d}{=}\left(\eta+\sum_{j=1}^{k+1}\tau_jx_j\right)^{-1},
    \label{eq:analytic_recursion}
\end{equation}
where $\tau=t^2$ and all incoming fields and hoppings are independent. The symbol $\overset{d}{=}$ indicates equality in distribution.
The branching number is $k\geq2$: a cavity site has $k$ incoming branches, while a bulk site has $k+1$. We assume a nondegenerate stationary distribution as $\eta\to0^+$ and take this limit before examining its tails.

Let the squared-hopping density behave as $p_\tau(\tau)\sim B\tau^\lambda$ at $\tau\to0^+$, with $\lambda >-1$, which is satisfied by the hopping distribution in Eq.~\eqref{eq:p_t} (with $\lambda = -(1+a)/2$).
If $P_c(x)$ denotes the cavity-field density, one branch gives a contribution $w=\tau x$, whose density is
\begin{equation}
    \begin{aligned}
        P_w(w)&=\int_0^\infty\frac{dx}{x}\,P_c(x)p_\tau(w/x)\\
        &\sim B\langle x^{-1-\lambda}\rangle_c\,w^\lambda,
        \qquad w\to0^+.
    \end{aligned}
    \label{eq:analytic_branch}
\end{equation}
The required negative moment will be checked below. For the sum of $m$ positive independent branch contributions, $S_m=\sum_{j=1}^m w_j$, all terms must be small when $S_m$ is small. Convolution therefore gives
\begin{equation}
    P_{S_m}(S)\propto S^{m(1+\lambda)-1},\qquad S\to0^+.
    \label{eq:analytic_small_sum}
\end{equation}
Since $x\overset{d}{=}S_k^{-1}$, the change of variables
$P_c(x)=x^{-2}P_{S_k}(1/x)$ yields
\begin{equation}
    P_c(x)\propto x^{-1-\beta_c},\qquad
    \beta_c=k(1+\lambda),\qquad x\to\infty.
    \label{eq:analytic_cavity_tail}
\end{equation}

The opposite tail follows from a large denominator. Provided the
large-$\tau$ tail decays sufficiently rapidly, multiplying by $\tau$
preserves the upper-tail exponent of $x$:
$P_w(w)\propto w^{-1-\beta_c}$ at large $w$. A large sum of a fixed
number of variables with this power-law tail is dominated by one large
term. Thus $P_{S_m}(S)\propto S^{-1-\beta_c}$ at large $S$, independently
of $m$. Inverting $S_k$ now gives
\begin{equation}
    P_c(x)\propto x^{\beta_c-1},\qquad x\to0^+.
    \label{eq:analytic_cavity_lower}
\end{equation}
This also verifies the moment in Eq.~\eqref{eq:analytic_branch}:
its convergence at zero requires $\beta_c>1+\lambda$, which holds because
$\beta_c-(1+\lambda)=(k-1)(1+\lambda)>0$.

For the physical LDoS, the same argument uses $m=k+1$. The extra branch changes the upper-tail exponent, whereas the lower tail still comes from one large cavity field. We thus get
\begin{equation}
    P(\rho)\propto
    \begin{cases}
        \rho^\nu, & \rho\to0^+,\\
        \rho^{-1-\beta}, & \rho\to\infty,
    \end{cases}
    \qquad
    \begin{aligned}
        \nu&=k(1+\lambda)-1,\\
        \beta&=(k+1)(1+\lambda).
    \end{aligned}
    \label{eq:analytic_ldos_exponents}
\end{equation}
Writing the lower exponent as $\nu=1+\beta-\gamma$ gives
$\gamma=3+\lambda$, which relates the two asymptotic tails. While this matching does not establish the full MF inversion identity in Eq.~\eqref{sym}, it does not depend on $k$ and therefore does not depend on the geometry; it relies only on the tree structure of the cavity equations and the distribution of hoppings. A similar result was found in Ref.~\cite{cugliandolo2024}, where the exponent $\gamma_{\mathrm{ER}}=3$ does not depend on the average degree $p$ of the Erdös-Rényi random graphs.

For $p(t)\propto e^{-t^2}|t|^{-a}$, with $a<1$, changing variables to
$\tau=t^2$ gives $p_\tau(\tau)\propto e^{-\tau}\tau^{-(1+a)/2}$,
so $\lambda=-(1+a)/2$. For the branching number $k=2$ considered here,
\begin{equation}
    \begin{aligned}
        \beta_c&=1-a, & \nu&=-a,\\
        \beta&=\frac{3(1-a)}{2}, & \gamma&=\frac{5-a}{2}.
    \end{aligned}
    \label{eq:analytic_model_exponents}
\end{equation}

The derivation therefore uses the purely imaginary Green's function at $E=0$, enforced by the $E\leftrightarrow-E$ symmetry, and the cavity equation, which expresses its inverse as a sum of independent positive branch contributions.
The hopping distribution enters through its behavior near zero, $p(t)\sim C|t|^{-a}$ with $a<1$: it may vanish, diverge, or approach a finite nonzero value. In particular, purely Gaussian hoppings correspond to $a=0$, for which Eq.~\eqref{eq:analytic_model_exponents} gives $\beta=3/2$ and $\nu=0$. Thus, $P(\rho)\propto\rho^{-5/2}$ at large $\rho$, while it approaches a constant at small $\rho$. A vanishing or divergent hopping density at the origin is therefore not required for the LDoS power-law tail. An essential suppression or singularity of $p(t)$ at the origin is outside this derivation.
The Gaussian cutoff at large $|t|$ is not needed, provided the hopping distribution decays sufficiently rapidly for the moments used above to exist.
The derivation assumes a stationary distribution with power-law tails; it does not establish its existence or stability.

\section{Population dynamics numerics}
\label{sec:population_dyn}

\begin{figure}[tbh]
\centering
\includegraphics[width=0.8\linewidth]{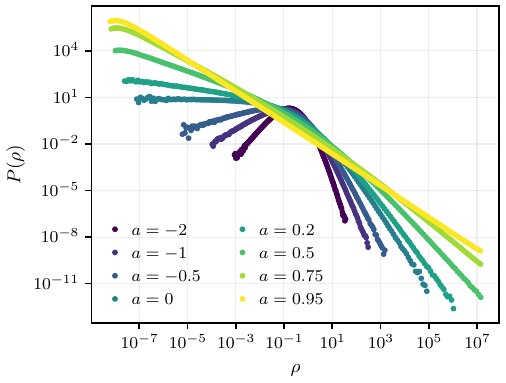}
\caption{Distribution function of the LDoS $P(\rho)$ at different values of $a$.}
\label{fig:P_rho_all}
\end{figure}
\begin{figure*}
\centering
\includegraphics[width=0.32\linewidth]{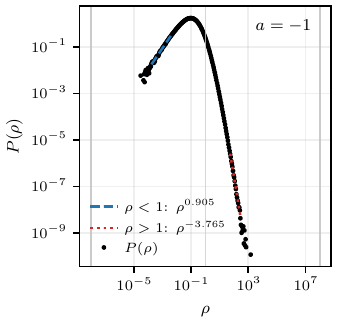}
\includegraphics[width=0.32\linewidth]{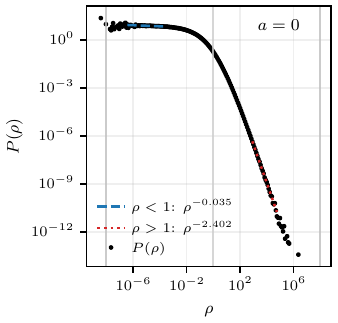}
\includegraphics[width=0.32\linewidth]{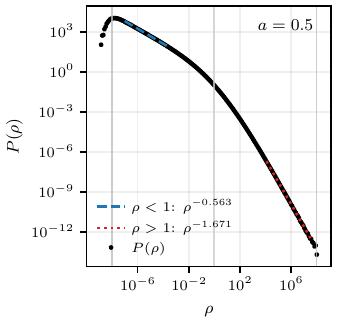}
\caption{LDoS distribution for $a=-1,\,0,\, 0.5$ (from left to right), with the power-law fits at $\rho\ll 1$ and $\rho \gg 1$ indicated in blue and red, respectively.}
\label{fig:P_rho}
\end{figure*}
A direct finite-size study at the band center is complicated by the boundary structure of a Cayley tree. For a tree of depth \(L\) and branching number \(k=2\), the two sublattices $A$ and $B$ contain different numbers of sites, \(\lvert N_A-N_B\rvert=2^L\). This imbalance produces at least \(2^L\) eigenstates at exactly zero energy~\cite{brouwer2002}. Since this number is extensive in the total system size, exact diagonalization at \(E=0\) is dominated by boundary-induced zero modes and does not provide a faithful probe of the bulk infinite-tree problem.

We therefore employ the cavity method, which works directly in the thermodynamic limit and avoids the boundary-induced sublattice imbalance. Here we describe its numerical implementation and compare the fitted tail exponents with the predictions of Sec.~\ref{sec:analytic_ldos}.
Writing \(z=E+i\eta\), the cavity Green's functions satisfy (the diagonal part entries are all vanishing for the chiral case studied here, $\epsilon_i=0$)
\begin{equation}
\label{eq:cavity_warmup}
    G_{i\to j}(z)= \left( z- \sum_{a\in\partial i\setminus j} t_{ia}^{2}G_{a\to i}(z) \right)^{-1}.
\end{equation}
We solve this recursion by population dynamics: the distribution of cavity fields is represented by a large population of complex numbers, which is iteratively updated by drawing \(k\) incoming fields and independent hopping amplitudes. After equilibration, diagonal Green's functions are reconstructed using the full coordination \(k+1\),
\begin{equation}
\label{eq:cavity_measure}
    G_{ii}(z)= \left( z- \sum_{j\in\partial i}t_{ij}^{2}G_{j\to i}(z) \right)^{-1}.
\end{equation}
We then sample the LDoS population \(\rho_i(E,\eta)= - \operatorname{Im}G_{ii}(E+i\eta)/\pi\) during the measurement sweeps and construct the corresponding distribution $P(\rho)$.

For the results presented in this work, we use a population of
\(n=10^6\) cavity fields. We perform \(5000\) warm-up sweeps, with each sweep consisting of \(n\) cavity updates according to
Eq.~\eqref{eq:cavity_warmup}. This is followed by \(8000\) measurement sweeps. During each measurement sweep, the cavity population is updated once and \(8000\) diagonal Green's functions are independently reconstructed according to Eq.~\eqref{eq:cavity_measure}. 
We have verified that our results are robust with respect to changes in the population size, the number of sweeps, and the random seed. We have also checked that the stationary distribution is independent of whether the initial population is purely imaginary or has a nonzero real part.

In Fig.~\ref{fig:P_rho_all}, we show the results of the population dynamics numerics for the distribution of LDoS $P(\rho)$. The main feature of the distribution at all values of $a$ in Eq.~\eqref{eq:p_t} is that we have two power-law segments at $\rho\ll 1$ and at $\rho\gg 1$.
The logarithmic-curvature analysis in the Supplemental Material~\cite{SupplementalMaterial} is consistent with power-law tails in the chiral model, whereas the apparent power-law regime of the nonchiral Anderson model becomes increasingly curved toward the tails.
The fit of the right tail of $P(\rho)$ to the power-law $P(\rho)\propto \rho^{-(1+\beta)}$ gives the value of the exponent $\beta$, see Fig.~\ref{fig:P_rho}. As it is shown in Sec.~\ref{sec:LDOS_distr} (Fig.~\ref{fig:beta_smaller_one}), $\beta$ is equal to the slope of the linear segment of the spectrum of fractal dimensions $f(\alpha)$. This slope also determines the exponent of the power-law tail in the distribution function $F(\ln(|\psi|^{2}))$: 
\begin{equation}\label{F_log_psi_2}
F(\ln(|\psi|^{2}))= C\,N^{f(\alpha)-1}\sim \frac{1}{|\psi|^{2\beta}}, \;\;\;\;\alpha=\frac{-\ln(|\psi|^{2})}{\ln N},
\end{equation}
where $C$ is the normalization constant.
 
\begin{figure}
\centering
\includegraphics[width=0.8\linewidth]{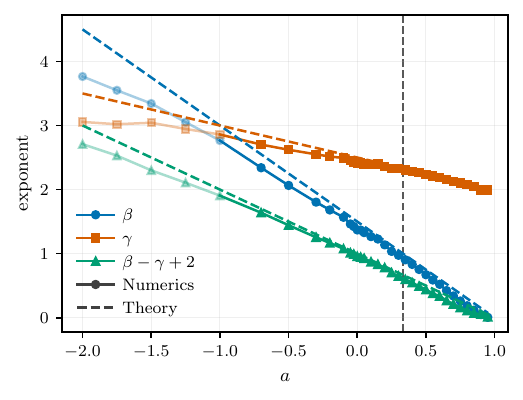}
 \caption{The fitted (solid) and analytically predicted (dashed) values of $\beta$, $\gamma$, and $\beta-\gamma+2$ vs. $a$. The vertical dashed line at $a=1/3$ marks the analytical prediction obtained by combining Eq.~\eqref{eq:analytic_model_exponents} with the criterion $\beta=1$. The numerical points at $a\leq-1$ are faded because their extraction is sensitive to the fitted tail range, as discussed in the Supplemental Material~\cite{SupplementalMaterial}.}
\label{fig:beta_gamma}
\end{figure}

The left-tail fit $P(\rho)\sim\rho^{\beta+1-\gamma}$ at $\rho\ll1$ (Fig.~\ref{fig:P_rho}), combined with the right-tail fit, gives $\gamma$. Figure~\ref{fig:beta_gamma} compares these finite-range estimates of $\beta$ and $\gamma$ with Eq.~\eqref{eq:analytic_model_exponents}. Their extraction becomes sensitive to the fitted tail range for $a\leq-1$, as we analyze in the Supplemental Material~\cite{SupplementalMaterial}.
Figure~\ref{fig:beta_gamma} demonstrates that at all values of $a<1$ studied we have $\beta\geq 0$, $\gamma\geq 2$ and $\beta\geq\gamma-2$. We remind here once again that in the chiral classes the symmetry parameter $\gamma$ is not fixed by the symmetry class but may depend on the details of the system \cite{gruzberg2011}. In our case, it depends on the parameter $a$ in the distribution of hopping strengths (see Eq.~\eqref{eq:p_t}). 

\begin{figure}
\centering
\includegraphics[width=0.8\linewidth]{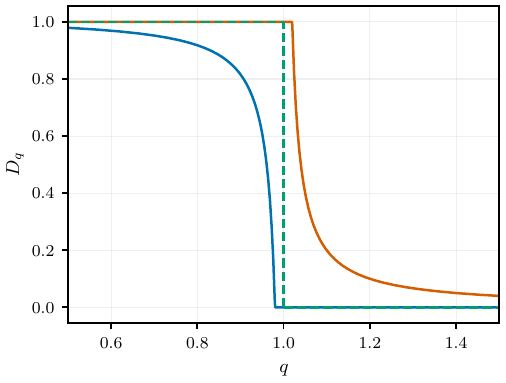}
\caption{$D_{q}$ for $\beta=0.98$ (blue) and $\beta=1.02$ (orange) and for the critical state (semi-localized) $\beta=1$ (green, dashed).}
\label{fig:critical}
\end{figure}

The criterion $\beta=1$ in Sec.~\ref{sec:LDOS_distr}, combined with Eq.~\eqref{eq:analytic_model_exponents}, gives the analytical prediction $a_c=1/3$. The finite-range fits in Fig.~\ref{fig:beta_gamma} instead give an estimate $a_c\approx0.25$. These values should be distinguished: their difference remains unresolved by the numerical results presented here. The Gaussian case $a=0$ lies in the semi-fractal regime according to both estimates. At the critical value $\beta=1$, a very unusual set of fractal dimensions $D_{q}$ is realized when $D_{q}=1$ for $q<1$ and $D_{q}=0$ for $q>1$ \cite{deluca2013support} [see Fig.~\ref{fig:critical}]. This dependence of  $D_{q}$ is an extreme form of semi-fractality: it is half-ergodic and half-localized.

\begin{figure}
\centering
\includegraphics[width=0.8\linewidth]{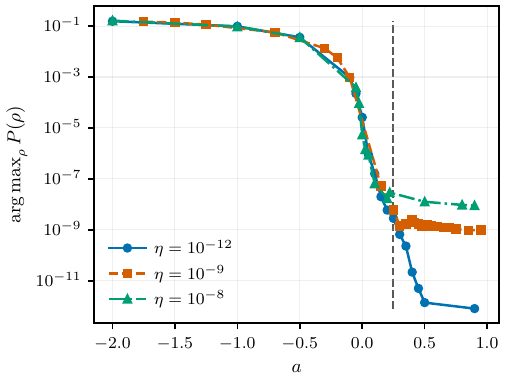}
\caption{Most probable value of $\rho$ vs. $a$. The vertical dashed line marks the numerical estimate $a\approx0.25$ obtained from the fitted upper-tail exponent $\beta=1$; it is distinct from the analytical prediction $a_c=1/3$.}
\label{fig:most_prob}
\end{figure}
The most probable value of $\rho$ decreases rapidly over the same range of $a$. Figures~\ref{fig:P_rho_all} and~\ref{fig:most_prob} show a drop from a value of order $0.1$ to $\sim\eta\sim10^{-12}$ over an interval $\Delta a\approx0.5$.

\section{Comparison between theory, population dynamics, and exact diagonalization numerics.}
\label{sec:theory_num}
\subsection{$P(\rho)$ from population dynamics and exact diagonalization}
\label{sec:PD_ED}
First of all, we compare the distribution of LDoS $P(\rho)$ computed by matrix inversion for a finite “chiral random regular graph" and that obtained by population dynamics. The “chiral RRG" is constructed as follows. We take two groups of $N/2$ points each and then connect each point of one group with three points of the other. There is no on-site disorder, and the strengths of the links are i.i.d. according to Eq.~\eqref{eq:p_t}. Such a graph is bipartite by construction, in contrast to the usual RRG with link disorder.
The measured distributions differ little between these two graph ensembles at the sizes $N=2^L$, $L=4,\dots,12$, considered here. We therefore use the usual RRG for the numerical results below. This agreement concerns the measured observables; it does not give the usual RRG an exact sublattice symmetry.
\begin{figure}
\centering
\includegraphics[width=0.8\linewidth]{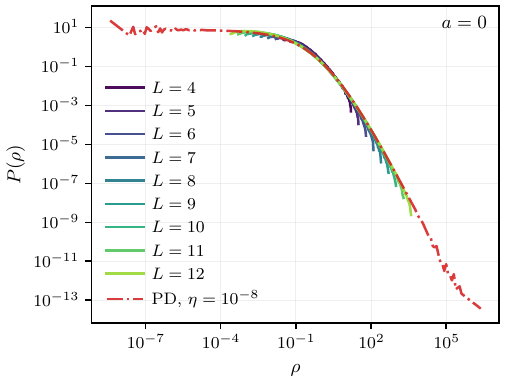}
\caption{Comparison of the LDoS distribution functions obtained in population dynamics (red dashed-dotted line) and by the exact diagonalization for $a=0$ and the sizes indicated, with $\eta = N^{-1} = 2^{-L}$.}
\label{fig:P_rho_compar}
\end{figure}
The result of the comparison in Fig.~\ref{fig:P_rho_compar} shows that $P(\rho)$ obtained by the exact diagonalization is quantitatively similar to that of population dynamics, although the power laws in the latter are much more accurate.

Note that for $a>a_{c}$, the coincidence of $P(\rho)$ computed by exact diagonalization with that computed in population dynamics depends on the value of $\eta$. The point is that the population dynamics corresponds to $N=\infty$ and thus to an infinite number of levels inside $\eta$. By contrast, the eigenfunction analysis in Sec.~\ref{sec:lnPsi} selects the state closest to $E=0$, and the conversion between LDoS and eigenfunction moments in Sec.~\ref{sec:model} uses a broadening containing one or a few levels. These are quite different limits in the case $\beta<1$ when the mean DoS is singular at $E=0$.

\begin{figure}
    \centering
    \includegraphics[ width=0.8\linewidth]{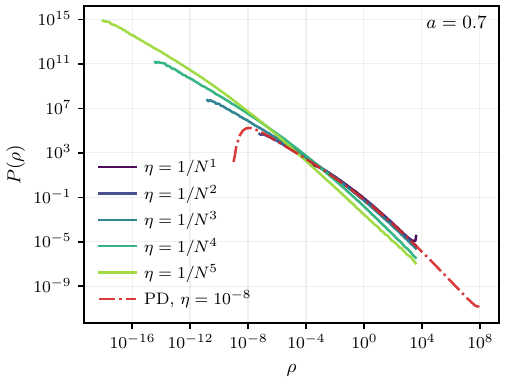}
    \caption{LDoS distribution for the RRG with $N=2^{12}$, $a=0.7$, and different $\eta$ (solid lines), compared with population dynamics on the Cayley tree (red dash-dotted line). The departure of the fitted upper-tail exponent from the population-dynamics value begins between $\eta=N^{-2}$ and $N^{-3}$, consistently with $\eta_c\sim N^{-1/\beta}$ and the analytical prediction $\beta=0.45$. At the smaller broadening $\eta\sim10^{-13}$, the fitted $\beta$ increases by approximately $0.1$.}
    \label{fig:eta_dep_of_beta}
\end{figure}

In Fig.~\ref{fig:eta_dep_of_beta}, the fitted upper-tail exponent begins to depart from the population-dynamics value as $\eta$ decreases from $N^{-2}$ to $N^{-3}$. This onset is compatible with the scaling estimate $\eta_c\sim N^{-1/\beta}$: the analytical value $\beta=0.45$ at $a=0.7$ gives $\eta_c\sim N^{-20/9}\sim10^{-8}$ for $N=2^{12}$, between these two broadenings. At the smaller broadening $\eta\sim10^{-13}$, the fitted $\beta$ has increased by approximately $0.1$ relative to population dynamics.
\subsection{Exact diagonalization results for the distribution of $\ln|\psi|^{2}$ and their correspondence to theory}
\label{sec:lnPsi}
\begin{figure*}
\centering
\includegraphics[ width=0.325\linewidth]{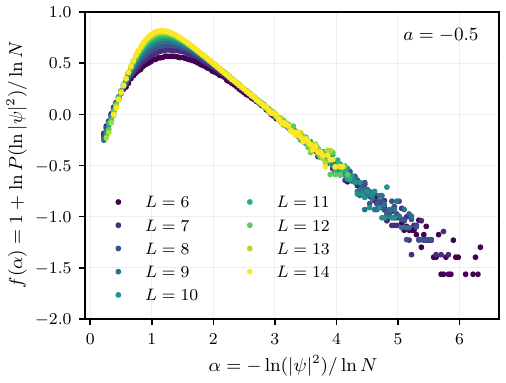}
\includegraphics[ width=0.32\linewidth]{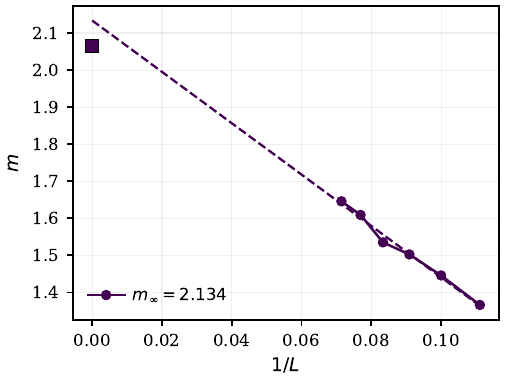}
\includegraphics[ width=0.32\linewidth]{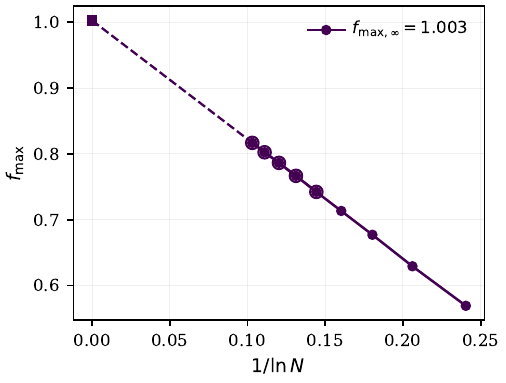}
\caption{{\it Left Panel}: $f(\alpha)$ for $a=-0.5$;  {\it Central panel}: the dependence of initial slope on $L$ and the $1/L$ extrapolation. The extrapolated value $\approx2.13$ is close to the fitted population-dynamics value $\beta\approx2.07$ (filled square); both are below the analytical prediction $\beta=2.25$ in Eq.~\eqref{eq:analytic_model_exponents};} {\it Right panel}: $1/L$ extrapolation of the maximal value of $f(\alpha)$.
\label{fig:f_a_alpha_m_0_5}
\end{figure*}
\begin{figure*}
\centering
\includegraphics[width=0.325\linewidth]{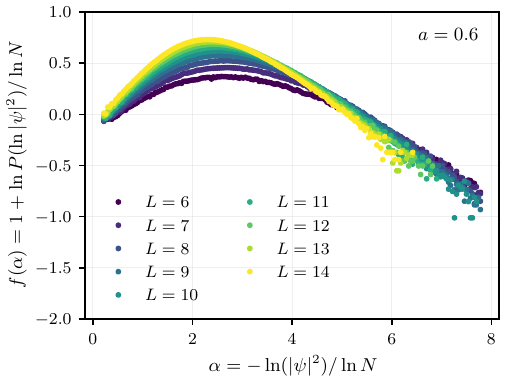}
\includegraphics[width=0.32\linewidth]{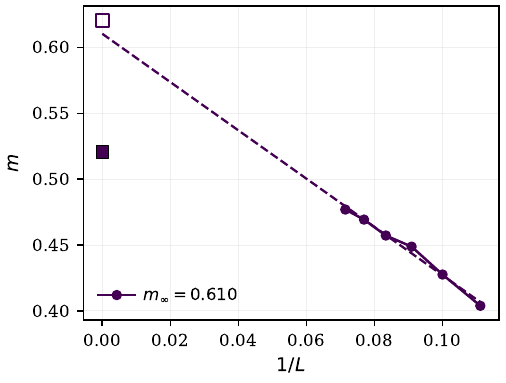}
\includegraphics[width=0.32\linewidth]{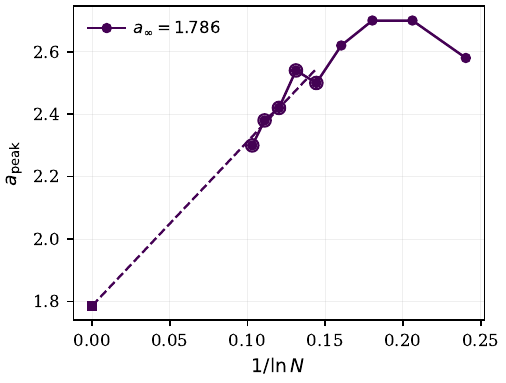}
\caption{{\it Left Panel}: $f(\alpha)$ for $a=0.6$;  {\it Central panel}: the dependence of initial slope on $L$ and the $1/L$ extrapolation. The extrapolated value $\approx0.61$ is close to the fitted finite-size value $\beta_{\mathrm{ED}}\approx\beta_{\mathrm{PD}}+0.1\approx0.62$ (empty square), where $\beta_{\mathrm{PD}}$ is the population-dynamics fit (filled square). The stationary analytical prediction is instead $\beta=0.6$; {\it Right panel}: $1/L$ extrapolation of the position of maximal value of $f(\alpha)$, giving an extrapolated peak position $\approx1.786$, compared with the fitted target $1/\beta_{\mathrm{ED}}\approx1.62$ and the analytical target $1/\beta=5/3 \approx 1.67$.}
\label{fig:f_a_alpha_0_6}
\end{figure*}
Now we check numerically the correspondence of $f(\alpha)$ to the theory of Sec.~\ref{sec:LDOS_distr}. We do it by exact diagonalization of the Anderson model on the RRG with i.i.d. link strengths determined by Eq.~\eqref{eq:p_t}. By collecting statistics of $|\psi|^{2}$ corresponding to the level with the minimal $|E_{n}|$ and finding the distribution function of $\ln|\psi|^{2}$, one can infer information about the spectrum of fractal dimensions $f(\alpha)$. The correspondence of this distribution to $f(\alpha)$ is given by Eq.~\eqref{F_log_psi_2}.

The examples of the distributions $F(\ln|\psi|^{2})$ and corresponding $f(\alpha)$ are shown below. We start by studying the case $\beta>1$.  
In the left panel of Fig.~\ref{fig:f_a_alpha_m_0_5} we present the plot of $f(\alpha)=1+\ln[F(\ln|\psi|^{2})]/\ln N$ vs $\alpha=-\ln|\psi|^{2}/\ln N$ which in the limit $L=\ln N/\ln 2\rightarrow \infty$ should represent the spectrum of fractal dimensions $f(\alpha)$. In this limit, the theory of Sec.~\ref{sec:LDOS_distr} predicts that the slope of the linear segment at $\alpha<1$ should be equal to $\beta$, and the maximum of the plot should be at $\alpha=1$ and should be equal to 1. The central and right panels of Fig.~\ref{fig:f_a_alpha_m_0_5} show the $1/L$ extrapolations. 
The extrapolated slope $\approx2.13$ lies between the fitted population-dynamics value $\beta\approx2.07$ and the analytical prediction $\beta=2.25$. Note that the nodes of the wave function heavily influence the part of the plot corresponding to $\alpha>1$ due to fast De Broglie oscillations and reflect the statistics of the nodes rather than the envelope of $\psi^{2}$, which (by definition) determines the spectrum of fractal dimensions.
It is only after rectifying the effect of the nodes that this part will be in quantitative correspondence with $f(\alpha)$ depicted in the inset of Fig.~\ref{fig:beta_smaller_one}.

Now we turn to a more difficult case $\beta<1$, reported in Fig.~\ref{fig:f_a_alpha_0_6}. The difficulty in this case is related to the singularity of the mean DoS and LDoS at $E=0$. For this finite-size comparison, we use the exponent fitted to the exact-diagonalization LDoS in the small-$\eta$ regime, rather than the population-dynamics fit, as discussed around Fig.~\ref{fig:eta_dep_of_beta}. The values $\beta_{\mathrm{ED}}\approx0.62$ and the extrapolated slope $\approx0.61$ are also close to the stationary analytical prediction $\beta=0.6$. The extrapolated peak position $\approx1.786$ lies above both the fitted target $1/\beta_{\mathrm{ED}}\approx1.62$ and the analytical target $5/3$.

\subsection{Power-law distributed eigenstate probabilities}
\label{sec:powerlaw-eigenstates}

The nonchiral deformed GUE of Ref.~\cite{truong2016} provides an example in which a deterministic power-law profile generates semi-fractality. Our model instead has i.i.d. hopping amplitudes and no imposed site-dependent variance profile. We ask whether a hierarchy of weights also emerges in its eigenstates near $E=0$. We present finite-size evidence for a power law in the disorder-averaged rank-ordered probabilities and examine its relation to the typical participation moments.

\begin{figure}
\centering
\includegraphics[width=0.8\linewidth]{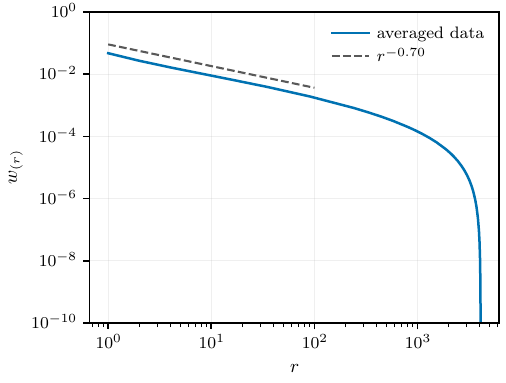}
\includegraphics[width=0.8\linewidth]{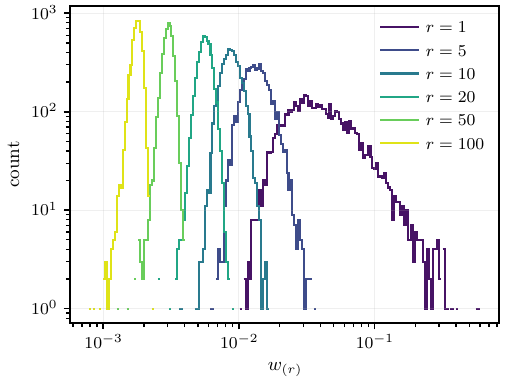}
\caption{\textit{Top panel}: Rank-ordered eigenstate probabilities, averaged over several eigenstates near $E=0$ and over $100$ disorder realizations in RRGs of size $N=2^{12}$ with $a=-0.5$. The rank $r$ denotes the position of a probability after the eigenstate weights have been sorted in decreasing order. The dashed lines are guides to the eye, with slopes obtained from log--log fits over $1\leq r\leq100$. 
\textit{Bottom panel}: Distribution of the eigenstate probabilities for fixed rank $r$, collected from $10$ eigenstates near $E=0$ and over $500$ disorder realizations, for $a=-0.5$. The measured fixed-rank distributions are relatively narrow for $r>1$, while $w_{(1)}$ is broadly distributed.
}
\label{fig:psi_sorted}
\end{figure}

We consider the RRG model introduced in Sec.~\ref{sec:PD_ED}. For each eigenstate, we order the probabilities
\begin{equation}
    w_i = |\langle i|\psi\rangle|^2
\end{equation}
in decreasing magnitude,
\begin{equation}
    w_{(1)} \geq w_{(2)} \geq \cdots \geq w_{(N)},
\end{equation}
and average the resulting rank-ordered profiles over a few eigenstates near $E=0$ and over disorder realizations. Figure~\ref{fig:psi_sorted} (top panel) shows a broad power-law regime in rank. To relate this observation to participation moments, we assume a profile whose normalization gives the size dependence
\begin{equation}
    \langle w_{(r)} \rangle
    \sim \frac{N^{-(1-\mu)}}{r^\mu},
    \qquad 0<\mu<1.
    \label{eq:sorted_weight_profile}
\end{equation}

The fixed-rank distributions in Fig.~\ref{fig:psi_sorted} (bottom panel) are relatively narrow for the measured ranks $r>1$, with no resolved power-law tails. If the distributions of $w_{(r)}/\langle w_{(r)}\rangle$ approach size-independent forms with finite moments of the orders considered, then, for fixed $q$,
\begin{equation}
    \langle w_{(r)}^q\rangle
    \sim \langle w_{(r)}\rangle^q,
    \qquad r>1,
    \label{eq:narrow_rank_distribution}
\end{equation}
up to an $N$-independent factor. Under this assumption, the mean rank-ordered profile determines the scaling of the participation moments of the power-law background. The fixed-size distributions shown here do not establish this assumption asymptotically.

Substituting Eq.~\eqref{eq:sorted_weight_profile} into
\begin{equation}
    I_q = \sum_{r=1}^{N}\langle w_{(r)}^q\rangle
\end{equation}
gives, for the power-law background,
\begin{equation}
    I_q^{\mathrm{bg}}
    \sim
    N^{-q(1-\mu)}
    \sum_{r} r^{-\mu q}.
\end{equation}
Consequently,
\begin{equation}
\label{eq:Pq}
    I_q^{\mathrm{bg}}
    \sim
    \begin{cases}
        N^{1-q}, & q<1/\mu,\\[3pt]
        N^{-q(1-\mu)}, & q>1/\mu.
    \end{cases}
\end{equation}
Equation~\eqref{eq:Pq} describes the background contribution and corresponds to the typical participation spectrum if the maximal weight follows the same scaling. The negative-$f(\alpha)$ part of the disorder-averaged spectrum describes rare events whose expected multiplicity, $N^{f(\alpha)}$, vanishes in a typical realization. The typical spectrum therefore terminates at $\alpha=\alpha_*$, where $f(\alpha_*)=0$.

Consistency with the linear segment $f(\alpha)=1+\beta(\alpha-1)$ in Sec.~\ref{sec:beta>1} requires $\mu=1/\beta$, giving $\alpha_*=1-1/\beta$. The resulting dimensions are $D_q^{\mathrm{typ}}=1$ for $q<\beta$ and $D_q^{\mathrm{typ}}=q(1-1/\beta)/(q-1)$ for $q>\beta$, as shown by the dashed curve in Fig.~\ref{fig:beta_smaller_one}.

By contrast, the disorder-averaged result in Eq.~\eqref{D_q} includes the negative-$f(\alpha)$ branch and gives $D_q=(\beta-1)/(q-1)$ for $q>\beta$. Thus, the rank hierarchy accounts for the typical spectrum under these assumptions, while recovering the mean moments also requires the statistics of rare large weights.

The maximal weight $w_{(1)}$ requires separate consideration because its distribution is much broader than those of the weights at $r>1$.
Consequently, its contribution to the participation moments must be computed from $\langle w_{(1)}^q\rangle$; in general, it cannot be inferred from $\langle w_{(1)}\rangle^q$. 
If the typical maximal weight scales as $N^{-(1-\mu)}$, it follows the power-law background and gives the typical moments in Eq.~\eqref{eq:Pq}. Its decrease in a typical realization does not exclude rare order-one weights from dominating disorder-averaged high moments.

A different way to obtain a semi-localized spectrum is for $w_{(1)}$ to remain of order unity as $N\to\infty$, while an extensive background retains a nonzero fraction of the norm and follows Eq.~\eqref{eq:sorted_weight_profile} with $0<\mu<1$. The peak then dominates $I_q$ for $q>1$, whereas the background dominates for $q<1$, giving
\begin{equation}
    \tau(q) =
    \begin{cases}
        q-1, & q<1,\\[3pt]
        0,   & q>1,
    \end{cases}
\end{equation}
This reproduces the fixed-$q\ne1$ spectrum of semi-localization.

\section{Discussion and Conclusions}
\label{sec:discussion}

We have derived two power-law tails of the band-center LDoS distribution for a particle hopping on a chiral Cayley tree with i.i.d. link amplitudes. The stationary cavity equations give $P(\rho)\propto\rho^{-a}$ at small $\rho$ and $P(\rho)\propto\rho^{-1-\beta}$ at large $\rho$, with $\beta=3(1-a)/2$. The exponent relating the two tails is $\gamma=(5-a)/2$. It varies continuously with the hopping-distribution parameter $a$, although the Hamiltonian remains in symmetry class BDI.

The origin of these tails is the combination of the $E\leftrightarrow-E$ pairing enforced by chirality and the structure of the cavity equations. The paired states have equal local weights, so the diagonal and cavity Green's functions are purely imaginary at $E=0$. As a consequence, the cavity equations reduce to inverse sums of independent positive branch contributions. Under the stationary-distribution and hopping-tail assumptions of Sec.~\ref{sec:analytic_ldos}, this combination is a sufficient condition for the two power-law tails and, when $\beta>1$, for semi-fractality. This condition includes Gaussian hoppings, for which $a=0$ and $\beta=3/2$; a singular hopping density at the origin is therefore not required. At finite observation energy, the band-center argument no longer applies, although the Hamiltonian remains chiral; the resulting crossover and the effect of on-site disorder are discussed in the Supplemental Material~\cite{SupplementalMaterial}.

For $\beta>1$, the LDoS moments yield an extensive eigenfunction support, with $D_1=1$, while the disorder-averaged higher moments give $D_q=(\beta-1)/(q-1)<1$ for $q>\beta$. As $\beta\to1^+$, the spectrum approaches the semi-localized limit: $D_q=1$ for $0\leq q<1$ and $D_q=0$ for $q>1$.
For $\beta<1$, the moments instead describe localization.

Population dynamics also provides evidence for the Mirlin-Fyodorov inversion relation, Eq.~\eqref{sym}, with an exponent $\gamma$ that depends on $a$. Such dependence in chiral classes is allowed by the field-theory analysis of Ref.~\cite{gruzberg2011}. 
Let us emphasize that the analytical matching of the two tails predicts $\gamma$, but does not prove the inversion relation for the full distribution.
For the problem on the tree, the cavity equations provide the simplest way to check both statements: population dynamics gives the full LDoS distribution directly, while the tail analysis predicts its exponents without finite-size diagonalization. Figure~\ref{fig:MF_symmetry} tests the inversion relation for Gaussian hoppings using a fitted $\gamma$, and Fig.~\ref{fig:beta_gamma} compares the fitted exponents with their analytical predictions.

Exact diagonalization gives a complementary description in terms of a power-law hierarchy of rank-ordered eigenstate weights. Under the scaling assumptions in Sec.~\ref{sec:powerlaw-eigenstates}, this hierarchy reproduces the typical participation spectrum; rare large weights must also be accounted for to recover the disorder-averaged high moments. The finite RRG calculations support this interpretation for the measured observables, without establishing chirality as the cause of the rank profile. A related hierarchy occurs in the momentum-space ground state of the long-range hopping model of Ref.~\cite{Bahovadinov2026}, where the weights are naturally ordered by momentum. In that model, an order-one weight at $p=0$ coexists with an extended background and produces semi-localization for $1<s<3/2$. Our finite-size rank profiles do not establish such a peak at the Cayley-tree transition.

The examples discussed in the Introduction suggest connections to other systems with exact or approximate chiral structure. In the tridiagonal representation of the $\beta_{\mathrm{RM}}$-ensemble, the diagonal disorder vanishes as $\beta_{\mathrm{RM}}\to\infty$, leaving a chiral nearest-neighbor chain~\cite{dumitriu2002,das2025}. Tree-like clusters near the percolation transition provide an approximate chiral structure in the Erd\H{o}s--R\'enyi problem~\cite{cugliandolo2024}, and we argue that chiral-symmetry-breaking terms are small in the experimental XY model~\cite{lunkin_feigelman_kravtsov26}. The nonchiral deformed GUE of Ref.~\cite{truong2016} provides a distinct mechanism and shows that chirality is not necessary for semi-fractality.

A direct extension is to formulate and solve cavity equations for tree models in other symmetry classes, including C, CI, and D. Do their LDoS distributions develop two power-law tails? Does a Mirlin-Fyodorov inversion relation hold, and what fixes $\gamma$? For C and CI, the field-theory values $\gamma=5$ and $\gamma=4$, respectively, provide predictions to test~\cite{gruzberg2011}; for class D, the cavity analysis must determine which inversion relation, if any, holds in the chosen model and what fixes its exponent. The corresponding cavity recursions would determine whether the positive scalar mechanism used here survives or must be replaced by a different one. We leave these questions for future work.

\section*{Acknowledgements}

VEK is grateful to M. V. Feigel'man and I. M. Khaymovich for multiple illuminating discussions. The Authors are grateful to P. Nosov for useful comments and suggestions. BA, VEK, and CV also thank the International Institute of Physics in Natal, BR, for the kind hospitality during the completion of this work. The simulations presented in this article were performed on computational resources managed and supported by Princeton Research Computing.

\section*{Data availability}

Some of the numerical analysis and plotting code used in this work was developed with assistance from GPT 5.6. The authors designed the algorithms, reviewed and modified the generated code, and verified the numerical outputs. The code is made available in the GitHub repository in Ref.~\cite{GitHub}.

\bibliography{references.bib}

\clearpage
\onecolumngrid
\setcounter{section}{0}
\setcounter{equation}{0}
\setcounter{figure}{0}
\setcounter{table}{0}

\renewcommand{\thesection}{S\Roman{section}}
\renewcommand{\theequation}{S\arabic{equation}}
\renewcommand{\thefigure}{S\arabic{figure}}
\renewcommand{\thetable}{S\Roman{table}}

\renewcommand{\theHsection}{supp.\arabic{section}}
\renewcommand{\theHequation}{supp.\arabic{equation}}
\renewcommand{\theHfigure}{supp.\arabic{figure}}
\renewcommand{\theHtable}{supp.\arabic{table}}

\section*{---Supplemental Material---\\
Semi-fractality and localization on a chiral Cayley tree}
\section{Effects of finite energy and on-site disorder}
\label{sec:sm_departure_chiral}

The power-law LDoS distribution studied in the main text is obtained at
$E=0$ without on-site disorder. At finite observation energy, the Green
function is evaluated away from the band center, but the random-hopping
Hamiltonian still has chiral symmetry. Adding independent on-site
energies $\epsilon_i\in[-W/2,W/2]$ breaks the chiral symmetry of the
Hamiltonian itself.

\begin{figure*}[!htb]
    \centering
    \includegraphics[width=\textwidth]{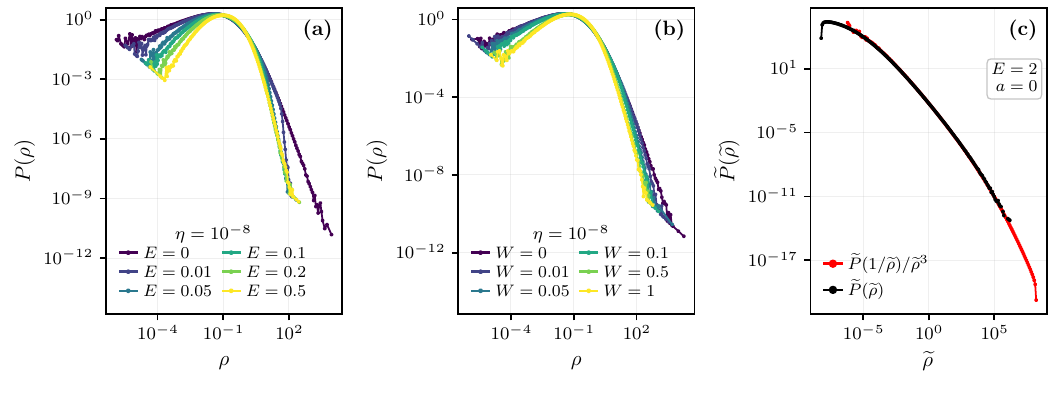}
    \caption{Effects of finite energy and on-site disorder on the LDoS
    distribution.
    (a) LDoS distributions at $a=-0.5$, $W=0$, and different $E$.
    (b) Distributions at $a=-0.5$, $E=0$, and uniform on-site disorder
    of width $W$. Both panels use $\eta=10^{-8}$.
    (c) Comparison at $E=2$, $a=0$, between the distribution
    $\widetilde P(\widetilde\rho)$ (black) and its inverted form
    $\widetilde P(1/\widetilde\rho)/\widetilde\rho^{3}$ (red).
    }
    \label{fig:sm_departure}
\end{figure*}

For small $E$ or $W$, the LDoS distribution follows the chiral result
over an intermediate range of $\rho$. Deviations appear first in the
tails [Fig.~\ref{fig:sm_departure}(a,b)]. At finite energy, the
crossover in the large-$\rho$ tail occurs at
\begin{equation}
    \rho_E\sim\frac{1}{|E|}.
    \label{eq:sm_energy_crossover}
\end{equation}
To estimate this scale, we compare the energy detuning with the inverse
Green function at the chiral band center. At $E=W=0$,
$|G^R_{ii}|=\pi\rho$, so the denominator of a large-$\rho$ Green
function is of order $1/(\pi\rho)$. The detuning becomes relevant when
$|E|\pi\rho$ is of order unity. Equation~\eqref{eq:sm_energy_crossover}
therefore gives the crossover scale up to a numerical prefactor; it
does not impose an upper bound on the LDoS.

As $E$ or $W$ increases, the range of $\rho$ over which the distribution
follows the chiral result shrinks. The upper tail becomes steeper, and
fits over a finite range give larger effective values of $\beta$.
Even a small detuning can produce a relatively sharp crossover, as
shown for $E=0.01$ in Fig.~\ref{fig:sm_departure}(a).
Farther from the band center, the comparison with the inverted
distribution is consistent with the Dyson value $\gamma=3$.
Figure~\ref{fig:sm_departure}(c) shows this comparison at $E=2$, $a=0$,
with the normalization $\widetilde\rho=\rho/\rho_0$ used in the main text.

\clearpage
\section{Extraction of the LDoS tail exponents at negative $a$}
\label{sec:sm_tail_extraction}

We fit the two tails of the LDoS distribution separately, using
$P(\rho)\propto\rho^\mu$ at small $\rho$ and
$P(\rho)\propto\rho^{-1-\beta}$ at large $\rho$.
We then calculate $\gamma=1+\beta-\mu$ from the fitted exponents.
For $k=2$, the stationary cavity analysis predicts
\begin{equation}
    \beta_{\mathrm{th}}=\frac{3(1-a)}{2},\qquad
    \mu_{\mathrm{th}}=-a,\qquad
    \gamma_{\mathrm{th}}=\frac{5-a}{2}.
    \label{eq:sm_tail_theory}
\end{equation}
The fits do not impose these relations.

We use the stored $E=0$ histograms without on-site disorder. Each
histogram has population size $M=10^6$ and contains
$N_s=6.4\times10^7$ local samples. The retained fraction
$\varepsilon$ sets the fitting range separately for each tail.
Starting from each end of the saved histogram, we retain complete bins
until adding the next bin would make the cumulative count exceed
$\varepsilon N_s$. Empty bins within the selected range are included
in the fit. Decreasing $\varepsilon$ therefore restricts the fit to
more extreme events. For example, $\varepsilon=10^{-5}$ retains at
most $640$ samples per tail. This parameter is distinct from the
multifractal moment order $q$.

For the retained bins, we maximize the log-likelihood
\begin{equation}
    \ell(b)=\sum_i n_i\ln\!
    \frac{e^{-bz_i}-e^{-bz_{i+1}}}{1-e^{-bZ}},
    \label{eq:sm_tail_likelihood}
\end{equation}
where $n_i$ is the number of observations in $[z_i,z_{i+1}]$ within the saved range
$[0,Z]$. For the lower tail, $z=\ln(\rho_-/\rho)$ and $b=\mu+1$;
for the upper tail, $z=\ln(\rho/\rho_+)$ and $b=\beta$, with
$\rho_\pm$ the respective thresholds. The bin probabilities in this
likelihood are obtained by integrating the power law over each bin.
We require at least $100$ counts and three bins in each tail.
The error bars are calculated from the curvature of the log-likelihood
under the assumption of independent samples. They do not account for
correlations within the population dynamics or for systematic errors.

\begin{figure*}[!tp]
    \centering
    \includegraphics[width=\textwidth]{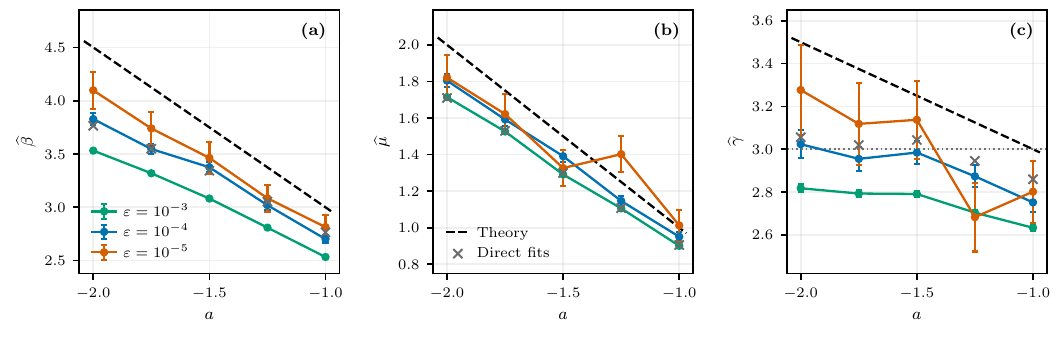}
    \caption{Effective LDoS exponents at $E=0$, $\eta=10^{-8}$, and
    $M=10^6$. Colors indicate the retained fraction $\varepsilon$;
    gray crosses show the earlier direct fits. Black dashed curves give
    Eq.~\eqref{eq:sm_tail_theory}; the dotted line marks $\gamma=3$.
    Error bars are calculated from the log-likelihood curvature assuming
    independent samples.}
    \label{fig:sm_tail_comparison}
\end{figure*}

\begin{figure*}
    \centering
    \includegraphics[width=\textwidth]{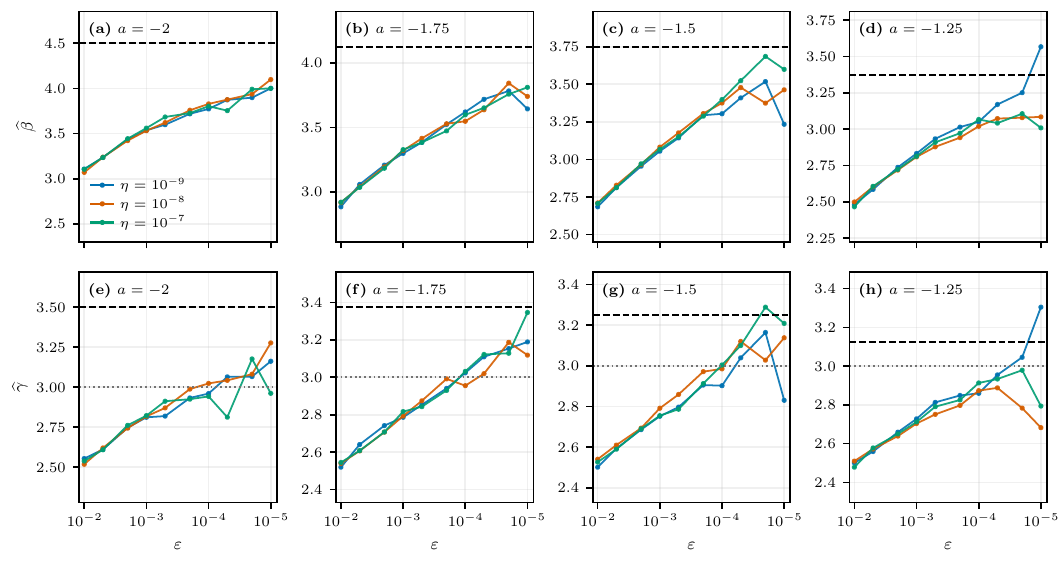}
    \caption{Threshold dependence of $\widehat\beta$ (top) and
    $\widehat\gamma$ (bottom), with $\varepsilon$ decreasing to the
    right. Colors distinguish $\eta$; one histogram per $\eta$ is shown
    for each $a$.
    Black dashed lines give the theoretical exponents; dotted lines
    mark $\gamma=3$.
    All populations have $M=10^6$ and at least $5000$ equilibration sweeps.}
    \label{fig:sm_tail_threshold_scan}
\end{figure*}

The fitted exponents depend on the selected range, as shown in
Figs.~\ref{fig:sm_tail_comparison} and~\ref{fig:sm_tail_threshold_scan}
and Table~\ref{tab:sm_tail_estimates}. At $a=-2$, reducing
$\varepsilon$ from $10^{-3}$ to $10^{-5}$ changes
$(\widehat\beta,\widehat\gamma)$ from $(3.532,2.818)$ to
$(4.098,3.277)$, closer to the theoretical values $(4.5,3.5)$.
The estimates continue to depend on the threshold when $\eta$ is varied.
There is also variation between runs: at $\varepsilon=10^{-5}$, six
distinct histograms with $a=-2$, $\eta=10^{-8}$ give $\widehat\gamma$
between $2.64$ and $3.30$. The available data do not establish whether
$\gamma$ saturates at $3$ or converges to Eq.~\eqref{eq:sm_tail_theory}
in the asymptotic tails.

\begin{table}[!htb]
    \caption{Tail exponents fitted at $\eta=10^{-8}$ and two retained
    fractions $\varepsilon$, using the primary histograms. The rows
    marked ``theory'' give Eq.~\eqref{eq:sm_tail_theory}.}
    \label{tab:sm_tail_estimates}
    \centering
    \begin{tabular}{rcrrr}
        \toprule
        $a$ & $\varepsilon$ & $\widehat\beta$ & $\widehat\mu$ & $\widehat\gamma$\\
        \midrule
-2 & $10^{-4}$ & 3.828 & 1.804 & 3.024 \\
-2 & $10^{-5}$ & 4.098 & 1.821 & 3.277 \\
-2 & theory & 4.500 & 2.000 & 3.500 \\
\addlinespace[3pt]
-1.75 & $10^{-4}$ & 3.547 & 1.592 & 2.955 \\
-1.75 & $10^{-5}$ & 3.740 & 1.621 & 3.119 \\
-1.75 & theory & 4.125 & 1.750 & 3.375 \\
\addlinespace[3pt]
-1.5 & $10^{-4}$ & 3.376 & 1.391 & 2.985 \\
-1.5 & $10^{-5}$ & 3.464 & 1.326 & 3.138 \\
-1.5 & theory & 3.750 & 1.500 & 3.250 \\
\addlinespace[3pt]
-1.25 & $10^{-4}$ & 3.020 & 1.146 & 2.874 \\
-1.25 & $10^{-5}$ & 3.085 & 1.402 & 2.683 \\
-1.25 & theory & 3.375 & 1.250 & 3.125 \\
        \bottomrule
    \end{tabular}
\end{table}

\section{Logarithmic curvature of the LDoS distribution}
\label{sec:sm_log_curvature}

\begin{figure*}
    \centering
    \includegraphics[width=0.49\textwidth]{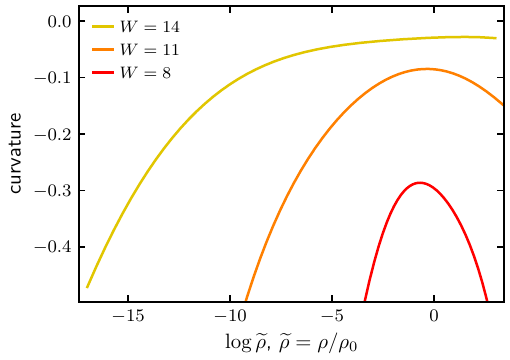}
    \hfill
    \includegraphics[width=0.49\textwidth]{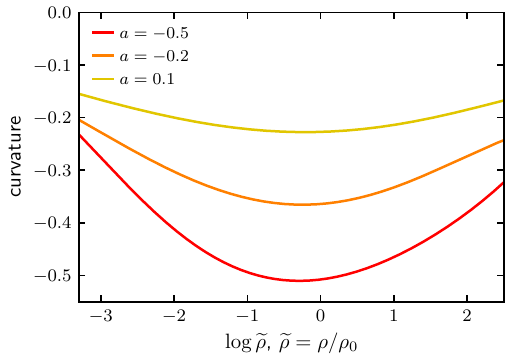}
    \caption{Logarithmic curvature
    $\kappa=d^2\ln\widetilde P/d(\ln\widetilde\rho)^2$ of the LDoS
    distribution at $E=0$, for branching number $k=2$ and population
    size $M=10^6$.
    Left: unit hopping and uniform on-site disorder of widths
    $W=8,11,14$.
    Right: random hopping with
    $p(t)\propto e^{-t^2}|t|^{-a}$, $a=-0.5,-0.2,0.1$, and no on-site
    disorder. A pure power law has zero logarithmic curvature.
    The scale $\rho_0=\langle\rho^q\rangle^{1/q}$ uses $q=1$ on the
    left and $q=0.6,0.5,0.4$, respectively, on the right.}
    \label{fig:sm_log_curvature}
\end{figure*}

An apparent power law over a finite interval can be distinguished from
the onset of a curved tail by examining the local logarithmic slope of
the LDoS distribution. We write $x=\ln\widetilde\rho$, where
$\widetilde\rho=\rho/\rho_0$ and
$\widetilde P(\widetilde\rho)=\rho_0P(\rho_0\widetilde\rho)$, and define
\begin{equation}
    s(x)=\frac{d\ln\widetilde P(e^x)}{dx},\qquad
    \kappa(x)=\frac{ds(x)}{dx}.
    \label{eq:sm_log_curvature}
\end{equation}
For $P(\rho)\propto\rho^p$, the slope is constant, $s=p$, and
$\kappa=0$. A nonzero curvature measures the change of the effective
exponent with $\rho$. Changing $\rho_0$ only shifts the horizontal
origin and does not change the shape of the curvature profile.

Figure~\ref{fig:sm_log_curvature} compares the nonchiral Anderson model
with unit hopping and uniform on-site disorder with the chiral
random-hopping model at $E=0$ and without on-site disorder. The two
curvature profiles have opposite shapes. In the chiral case, the
curvature has a negative minimum in the crossover region, and its
magnitude decreases on both sides. This behavior is consistent with a
crossover between two power-law tails, whose logarithmic slopes become
constant away from the crossover. For the stationary chiral solution
discussed in the main text, the predicted limits at $k=2$ are
\begin{equation}
    s(x)\longrightarrow
    \begin{cases}
        -a, & x\to-\infty,\\
        -1-\dfrac{3(1-a)}{2}, & x\to+\infty.
    \end{cases}
    \label{eq:sm_chiral_slope_limits}
\end{equation}
The curvature is therefore expected to vanish in both asymptotic
regimes, under the assumptions of the stationary cavity analysis.

For the nonchiral distributions, the curvature is least negative in
the central region and becomes more negative toward the tails over
the resolved range. The effective exponent thus changes more rapidly
as one moves away from the central, approximately power-law regime.
The distinction is particularly clear on the small-$\rho$ side, where
the departure from zero curvature grows over many logarithmic units.
These curves indicate that the apparent power law is spoiled toward
the tails in the nonchiral model, whereas the chiral curves are
consistent with the approach to the power-law tails predicted by the
cavity analysis.

To estimate $\kappa$, we first calculate adjacent finite differences of
$\ln P$ with respect to $\ln\rho$. We fit their dependence on $x$ with
a sixth-degree polynomial over a selected range of populated bins and
differentiate the fitted polynomial. The fits describe the resolved
range only and are not extrapolated to $x\to\pm\infty$. In particular,
the decrease of $|\kappa|$ in the chiral curves supports the power-law
derived and observed in the main text.

\end{document}